# Strain-assisted magnetization reversal in Co/Ni multilayers with perpendicular magnetic anisotropy

**Gopman, D. B.[1,*], Dennis, C. L.[1], Chen, P. J.[1], Iunin, Y. L.[1,2], P. Finkel[3], M. Staruch[3], and Shull, R. D.[1]**

[1]Materials Science and Engineering Division, NIST, Gaithersburg, MD, 20899, USA
[2]Institute of Solid State Physics, RAS, Chernogolovka, 142432, Russia
[3]U.S. Naval Research Laboratory, Washington, DC, 20375, USA
[*]daniel.gopman@nist.gov

## ABSTRACT

Multifunctional materials composed of ultrathin magnetic films with perpendicular magnetic anisotropy combined with ferroelectric substrates represent a new approach toward low power, fast, high density spintronics. Here we demonstrate Co/Ni multilayered films with tunable saturation magnetization and perpendicular anisotropy grown directly on ferroelectric PZT [Pb(Zr$_x$Ti$_{1-x}$)O$_3$] substrate plates. Electric fields up to ± 2 MV/m expand the PZT by 0.1% and generate at least 0.02% in-plane compression in the Co/Ni multilayered film. Modifying the strain with a voltage can reduce the coercive field by over 30%. We also demonstrate that alternating in-plane tensile and compressive strains (less than 0.01%) can be used to propagate magnetic domain walls. This ability to manipulate high anisotropy magnetic thin films could prove useful for lowering the switching energy for magnetic elements in future voltage-controlled spintronic devices.

## Introduction

Perpendicular magnetic anisotropy (PMA) continues to generate significant technological interest, particularly due to the superior thermal stability over in-plane anisotropies when scaling down materials for high density magnetic data storage and spintronics applications.[1,2] The non-volatility required in magnetic nano-objects for high-density storage and logic requires high anisotropy thin film materials such that the ratio of anisotropy energy to thermal energy ($KV/k_BT$) is greater than 50, where $K$ is the anisotropy energy density, $V$ is the volume of a single magnetic nano-object, $k_B$ is the Boltzmann constant and $T$ is the temperature.[3,4] At the same time, the energy consumed by switching the magnetization must be minimized in order to realize energy efficient magnetic random access memory (MRAM) devices. One way to reduce the energy required for reversing the magnetization while maintaining sufficient magnetic anisotropy energy for thermal stability is to independently control the magnetization by a variable other than an applied external magnetic field.[1]

Magnetoelectric coupling is a promising method for this independent control of magnetization. A voltage applied to a ferroelectric substrate generates strains that can mechanically couple to a ferromagnetic element and modify its magnetic anisotropy.[5] Previous studies have explored strain-induced changes in magnetic anisotropy energy and domain wall propagation in hybrid piezoelectric/ferromagnetic heterostructures, for both in-plane and perpendicularly magnetized ferromagnetic thin films either by direct deposition[6-10] of the magnetic film onto the piezoelectric element or by indirect bonding[11,12] of the piezoelectric element onto a magnetic film on substrate. Among the materials showing PMA, Co/Pd, Co/Pt and CoFeB have been explored previously, largely due to their promise for high density magnetic recording and MRAM.

Co/Ni multilayers have attracted interest for their use in spin-valve and domain wall devices.[13-16] Until now this system has not been explored as part of such a ferromagnetic/ferroelectric heterostructure. Composed entirely of transition-metal ferromagnets, Co/Ni multilayers exhibit a tunable saturation magnetization by alternating the Co to Ni ratio and exhibit high spin-polarization due to the relatively lower spin-orbit coupling in this system compared with other PMA systems (Co/Pt, Co/Pd, FePt).[17] Moreover, the magnetoelastic contributions to the total magnetic energy from uniaxial out-of-plane (biaxial in-plane) strain in bulk cubic (111) Co and Ni have opposite signs, which makes this a particularly interesting material system to study strain-induced changes in magnetism in ultrathin layers of cobalt and nickel.[18-20]

Here we demonstrate that Co/Ni multilayers can be grown directly on top of ferroelectric $Pb(Zr_xTi_{1-x})O_3$ (PZT) substrate plates. This is distinct from previous studies that bonded PZT transducers to ferromagnetic films grown on non-piezoelectric substrates, and allows direct strain coupling between overlaid film and substrate.[11,12] We tune the magnetization and the PMA energy by changing the relative layer thicknesses of Co and Ni and by varying the number of layer repeats. Using the Co/Ni multilayer showing the strongest PMA energy, we investigate strain-assisted magnetization reversal induced by voltages applied to the PZT substrate. We find that the voltages applied to the PZT couple directly to the Co/Ni multilayers and that reductions in the coercivity of the magnetic film track the PZT strain-voltage curve with a -7 mT / % strain dependence, permitting us to tune the coercivity by 30%, commensurate with achieved strain-mediated coercivity reductions in other candidate materials for voltage-assisted magnetization reversal with PMA. This large modification of the magnetic coercivity caused by interfacial strain coupling shows promise for voltage-assisted magnetization reversal in Co/Ni multilayers /PZT heterostructures.

## Results

**Perpendicular Magnetic Anisotropy of Co/Ni Multilayers on PZT**

The multilayered film structures used in this study are prepared on PZT substrates by dc magnetron sputtering at room temperature with a base pressure below 6.6 x $10^{-6}$ Pa (5 x $10^{-8}$ Torr). Co/Ni multilayers are grown on top of a seed layer of Ta(3 nm)/Pt(2 nm) and are capped by a Pt(1.6 nm)/Ta(3 nm) bilayer. For magnetization (*M)* versus applied field ($\mu_0 H$) and ferromagnetic resonance (FMR) measurements, five different multilayers are prepared: [Co(0.15)/Ni(0.3)]$_{x16}$; [Co(0.2)/Ni(0.4)]$_{x16}$/Co(0.2); [Co(0.2)/Ni(0.6)]$_{x16}$/Co(0.2); [Co(0.15)/Ni(0.6)]$_{x16}$/Co(0.15) and [Co(0.15)/Ni(0.6)]$_{x4}$/Co(0.15) (numbers in parentheses are nominal thicknesses in nm and subscripts indicate the number of bilayer repeats).

Figure 1 shows the in-plane and out-of-plane magnetization versus external magnetic field (*M-H*) curves for the [Co(0.15)/Ni(0.6)]$_{x16}$/Co(0.15) film, as determined from vibrating sample magnetometry. It can be seen that the sample shows a clear perpendicular easy axis when comparing the in-plane and out-of-plane magnetization loops. The saturation magnetization can be tuned by adjusting the relative thicknesses of Co and Ni, as well as the number of repeats, as can be seen in Figure 2(a). We obtain the largest saturation magnetization for the [Co(0.2)/Ni(0.4)]$_{x16}$/Co(0.2) film ($M_S$ = 780 ± 40 kA/m). With increasing nickel content relative to cobalt in the bilayer period, a decrease in the saturation magnetization is observed, a trend in agreement with the lower saturation magnetization of Ni relative to Co.

The perpendicular anisotropy field $\mu_0 H_{\text{eff}}$ was estimated from fits to the out-of-plane ferromagnetic resonance field versus frequency measured between 10 GHz and 25 GHz. The out-of-plane resonance field is linearly dependent on frequency, according to the Kittel equation[21]:

$$f = \mu_0\gamma(H_{res} + H_{eff}), \quad (1)$$

$$\mu_0 H_{eff} = 2\mu_0 \frac{K_1}{M_s} - \mu_0 M_s N_{zz}, \quad (2)$$

for which $f$ is the microwave excitation frequency, $\mu_0 H_{res}$ is the resonant field, and $\gamma$ is the gyromagnetic ratio. The effective magnetic anisotropy field ($\mu_0 H_{eff}$) comprises the surface and interface perpendicular anisotropy ($K_1$) as well as the demagnetization field ($\mu_0 M_s N_{zz}$), where the demagnetizing factor ($N_{zz}$) for a continuous thin film is equal to one. For $\mu_0 H_{eff} > 0$, the magnetization is oriented perpendicular to the film plane. Figure 2(b) presents $\mu_0 H_{eff}$ for the five multilayers, showing a net PMA for all samples except for the [Co(0.15)/Ni(0.3)]$_{x16}$/Co(0.15) film ($\mu_0 H_{eff}$ = -137 ± 1 mT). Decreasing the Co to Ni ratio reduces the demagnetization field ($\mu_0 M_s N_{zz}$) and can lead to gains in the PMA, particularly for the two Co(0.15)/Ni(0.6) multilayers. By reducing the number of Co/Ni bilayer repeats from sixteen to four, and correspondingly the overall magnetic thickness, the largest PMA ($\mu_0 H_{eff}$ = 268 ± 7 mT) is seen in the [Co(0.15)/Ni(0.6)]$_{x4}$/Co(0.15) film. This is likely a reflection of the relatively larger contribution (compared to total anisotropy) of the Pt/Co interfacial surface anisotropy generated in the seed and capping layers, which is known to be larger than the surface anisotropy at the Co/Ni interfaces.[13,22-24] Finally, the PMA energy $K_{eff}$ = ($\mu_0 H_{eff} M_s/2$) of multilayers was estimated from the anisotropy field $\mu_0 H_{eff}$ and the saturation magnetization $M_s$ and is displayed in Figure 2(c). For the highest PMA film, the PMA energy density $K_{eff}$ = (0.95 ± 0.08) x 10$^5$ J/m$^3$, a value high enough for non-volatile storage ($K_{eff}V/k_B T > 50$) in a 3 nm thick magnetic nano-element at reduced dimensions (30 nm diameter).

**Strain-assisted magnetization reversal**

The next step is to determine the effect of strain on magnetization reversal in the [Co(0.15)/Ni(0.6)]$_{x4}$/Co(0.15) multilayered film which had the largest PMA. Electrical contact was made to both the top and bottom of the 1 mm thick PZT substrate in order to apply voltages and strain the PZT. We used a silver epoxy to make a conductive bond between a copper plate and the bottom of the PZT substrate. Similarly, a 0.2 mm copper wire was bonded to the top of the multilayered film to apply voltages to the top surface of the PZT. We measured the effects of the applied voltages by characterizing the magnetic hysteresis loops using Kerr microscopy. Figure 3(a) shows out-of-plane hysteresis loops of the Co/Ni multilayers while applying electric fields of +1 MV/m, 0 and -2 MV/m to a 1 mm thick PZT substrate. The coercive field depends on the applied electric field magnitude and polarity – under a positive electric field (1 MV/m), the coercive field is maximal and then reaches a minimum for a negative electric field (-2 MV/m). In Figure 3(b) we present a non-linear magnetic coercivity trend under a series of applied electric fields beginning at -2 MV/m and then increasing in increments of +1 MV/m increments. We observe a trend in which the coercivity increases with increasing positive electric fields to a point, from $\mu_0 H_c$= 3.7 ± 0.1 mT for $E_{[001]}$ = -2 MV/m up to $\mu_0 H_c$= 4.6 ± 0.1 mT for $E_{[001]}$ = +1 MV/m, followed by a decrease with further increases in the electric field to $\mu_0 H_c$= 4.0 ± 0.1 mT for $E_{[001]}$ = +2 MV/m.

The non-linear relationship between the magnetic coercivity and the applied electric field is a direct consequence of the non-linear strain-voltage relationship in the PZT substrate. In Figure 3(c), we show the strain in PZT as a function of applied electric field. Strains in the PZT along the poling direction were determined by measured displacements in a homemade linear variable differential transformer. The curve for a full sweep of the electric field is a non-linear, butterfly-shaped loop, with a maximum tensile strain in the PZT in excess of 0.15%. To better estimate the electric field-strain relationship for arbitrary electric fields, we interpolated the PZT strain-voltage butterfly curve with third order polynomials to two sections of the strain curve (e.g. from -2 MV/m to +1 MV/m and from +1.5 MV/m to +2 MV/m). Using this parameterization on the measured magnetic coercivity versus electric field data, we find that the coercivity-voltage relationship for the Co/Ni multilayers (solid curves in Figure 3(b)) is linearly proportional to the PZT strain-voltage relationship, which is consistent with previous results for strain-induced modification of the PMA.[9] From the strain-coercivity relationship, we extrapolate that a 1% tensile strain of the PZT thickness would generate a 7 mT coercivity reduction.

To better understand the relationship between strain in the PZT and reduction in the Co/Ni multilayer coercivity, we estimate the magnetoelastic anisotropy change associated with a 0.15% elongation of the PZT substrate thickness (corresponding to a 1.0 +/- 0.2 mT coercivity reduction). The bulk magnetoelastic anisotropy change for a face-centered-cubic structure, where the out-of-plane magnetization aligns with the (111) direction, the change in magnetoelastic energy is estimated as

$$\Delta E_{elastic} = \frac{16 B_2^{Ni} + 5 B_2^{Co}}{21} (\Delta \varepsilon_\parallel - \Delta \varepsilon_\perp) \qquad (3)$$

where $B_2^{Co}$ = -29 MJ/m³ and $B_2^{Ni}$ = +10 MJ/m³ reflect the (111) cubic magneto-elastic coupling coefficients of Co and Ni.[25] If we then assume perfect transmission of in-plane strain from the PZT ($\nu_{PZT} = 0.44$), we arrive at $\Delta\varepsilon_\parallel = -6.6 \times 10^{-4}$, and subsequently the out-of-plane strain through elastic deformation of the Co/Ni $\Delta\varepsilon_\perp = -\nu\Delta\varepsilon_\parallel = 2.2 \times 10^{-4}$, where we attribute a Poisson ratio of $\nu = 0.33$ for the metallic Co/Ni multilayers. From these values, we estimate a change in magnetoelastic energy of -1.9 kJ/m³, or a reduction in the PMA field by 1.8 mT. Although this calculated value is 80% larger than our measured coercivity change, it lends additional support to the hypothesis that the coercivity is strain-mediated, but with room for further improvements. Furthermore, the calculated change in the PMA energy is less than a 1% reduction in the total PMA energy. This modest reduction in the PMA is expected based on the very large interfacial PMA relative to volume magnetoelastic anisotropy as has been reported previously.[20,26]

We have also examined the possibility of using voltage-induced strains in the PZT to depin domain walls and reverse the magnetization. Figure 4 shows two Kerr microscopy images taken under a modest applied out-of-plane magnetic field (-2.4 mT) below $\mu_0 H_c$(-4.2 mT) after applying a large, saturating positive magnetic field (28.6 mT) and alternating the direction of a 1 MV/m electric field in the PZT. Prior to applying an electric field, some reversed domains have already nucleated and the reduced magnetization is stable at a value of $M/M_s = 0.64 \pm 0.06$. However, following the application of a -1 MV/m electric field, we observe a significant increase in the area occupied by reversed magnetic domains seen in Fig. 4(a). This corresponds to a reduced magnetization of 0.12 +/ 0.06. Then, after reversing the polarity of the electric field to +1 MV/m, we show another moderate increase in reversed domains (Fig. 4(b)), which assists in reversing the net magnetization direction ($M/M_s$ = -0.10 ± 0.07).

We also note that the size of reversed domains increases under applied electric fields. As shown in Fig. 4(c), the average reversed domain size increases from 152 ± 4 µm² before applying an electric field to 676 ± 29 µm² after applying -1 MV/m and increasing to 806 ± 36 µm² after reversing the polarity to +1 MV/m. The observed bi-polar enhancement of domain expansion may reflect the inhomogeneous ferroelectric domain polarization, permitting local reductions *and* enhancements of the magnetic domain wall pinning energy due to opposite magnetoelastic energy contributions generated in regions of the Co/Ni film strained in proximity to ferroelectric "up" or "down" domains.[27] This demonstrates that domain expansion, as a consequence of domain wall de-pinning, can be enhanced by strain. This is in good agreement with previous studies that have demonstrated that strains in PMA films can enhance domain wall propagation by reducing the perpendicular anisotropy.[9,11]

**Lowering the reversal barrier with strain modification**

Finally, we demonstrate proof-of-concept for this system for electrically-assisted switching of the magnetization. In Figure 3(a) we saw that the magnetic coercivity of the Co/Ni multilayers depends on the electric field applied to the PZT. The coercivity can be reduced to $\mu_0 H_c$= 3.5 ± 0.1 mT for $E_{[001]}$ = -2 MV/m compared with $\mu_0 H_c$= 4.6 ± 0.1 mT for $E_{[001]}$ = +1 MV/m, a coercivity reduction of more than 30%. The electric-field modulation of the coercivity in the Co/Ni multilayers could be used to switch the magnetization.

Figure 5 illustrates a concept in which strain-assisted magnetization reversal might be implemented. An out-of-plane bias magnetic field ($\mu_0 H_b$=-4.2 mT) is applied to a Co/Ni multilayered film on a PZT substrate under +1 MV/m electric field bias after saturation in a large positive magnetic field. This can be seen as the red solid line extending from large positive field down to -4.2 mT and denoted by a yellow star. Changing the electric bias to -2 MV/m drives the Co/Ni film into the low coercivity operating region, indicated by the arrow extending down from the yellow star. The Co/Ni film would subsequently follow the hysteresis curve defined by the low coercivity state, represented by the blue curve. For device implementation, the strain would be modified locally with voltages applied to a contact region of the piezoelectric element directly adjacent or underneath the magnetic film or nanostructure. Furthermore, it would be desirable to transfer larger in-plane strains, which could be implemented by in-plane poling of a thin-film ferroelectric underlayer such as PMN-PT grown under the magnetic film.[9] In such a geometry, a modest increase of the in-plane strain to 0.3% would reduce the PMA energy of a 3 nm thick, 30 nm diameter nanomagnet by 5 $k_B T$ ($\Delta E_{elastic} \times V$), which would significantly reduce the energy cost to switch the magnetization. This could enable a non-volatile strain-based memory, in which the non-volatility of high PMA magnets is combined with the low power consumption of magnetoelectric control.

In summary, we have exploited the tunable perpendicular magnetic anisotropy in Co/Ni multilayered films and elastic coupling to a PZT substrate for the development of voltage-assisted magnetization reversal. This hybrid ferroelectric-ferromagnetic system exhibits strong PMA, sufficient for producing sub-30 nm diameter, ultrathin magnetic nano-objects with good thermal stability. The magnetoelectric coupling in this system that can reduce the Co/Ni coercive field is apparently strain mediated through lateral compression generated in the PZT, evidenced by the nearly identically non-linear strain-voltage and magnetic coercivity-voltage trends. We estimate that expansion of the PZT by 0.15% mediates a small reduction in the Co/Ni PMA energy due to increased magnetoelastic cost of the perpendicular magnetization under out-of-plane tensile strain in the Co/Ni multilayers. This PMA reduction was followed by a pronounced change in the Co/Ni coercive field of greater than 30%, which could potentially be used for voltage-assisted magnetization reversal applications. Finally, we proposed a concept for the implementation of a strain-mediated magnetization reversal, in which an applied voltage is used to lower a Co/Ni nanomagnet's coercivity below a moderate applied bias magnetic field. Therefore, high PMA Co/Ni multilayers combined with a PZT substrate are a promising materials system for voltage-controlled magnetic data storage and spintronics applications, in which the non-volatility of the PMA material is combined with the voltage-enabled reduction in switching energy.

## Methods

One-mm-thick $Pb(Zr_xTi_{1-x})O_3$ (PZT) plates were obtained from DeL Piezo Specialties, LLC (West Palm Beach, Florida, USA).[28] Typical surface roughness of at least 100 nm *rms* was reduced to 3 nm *rms* by using a chemical-mechanical polishing (CMP) tool (Bruker CP-4) at the NIST Nanofabrication Facility. The PZT plates were then processed using the following CMP recipe module: 5 minutes under 4 psi (28 kPa) with a 60 mL/minute flow of a 1:5 silica slurry (Eminess 556 Colloidal Silica) to distilled water dilution followed by a 2 minute rinse under 1 psi (7 kPa) with a 90 mL/minute flow of distilled water.[28] This cycle was repeated eight times, after which the PZT plate was promptly transferred from the CMP wafer holder to bath of a dilute ammonia solution heated to 80 degrees C for five minutes, followed by a distilled water soak with ultrasonic agitation for three minutes. This post-CMP process was critical to the removal of remaining slurry particles that would otherwise adhere to the PZT surface.

Plates were cleaved into substrate pieces using a semi-automated dicing saw. After spin-coating a thin layer of Shipley 1813 (MicroChem Corp., Westborough, Massachusetts, USA) photoresist onto the surface to protect the surface during dicing, 5 mm x 5 mm substrate pieces were cleaved from a polished PZT plate.[28] The pieces were subsequently cleaned using ultrasonic agitation in acetone for 5 minutes, followed by 5 minutes of ultrasonic agitation in isopropanol, and a rinse in distilled water, followed by drying with compressed air.

Magnetic multilayered films were grown on the substrate pieces using dc magnetron sputtering. Following a 5 minute clean under low energy $Ar^+$ ion bombardment (500 eV) to remove organic contamination, Co/Ni multilayered films of varying thicknesses were grown with identical seed layers, starting with 2 nm Pt on top of 3 nm Ta. Multilayered films all had the same capping sequence, with 3 nm Ta on top of 1.6 nm Pt. Five different Co/Ni multilayers were synthesized: $[Co(0.15)/Ni(0.3)]_{x16}/Co(0.15)$; $[Co(0.2)/Ni(0.4)]_{x16}/Co(0.2)$; $[Co(0.2)/Ni(0.6)]_{x16}/Co(0.2)$; $[Co(0.15)/Ni(0.6)]_{x16}/Co(0.15)$; $[Co(0.15)/Ni(0.6)]_{x4}/Co(0.15)$ where all thicknesses are given in nanometers. Deposition rates were calibrated using two parallel quartz crystal monitors.

Magnetic measurements were conducted using a vibrating sample magnetometer (VSM, MicroSense, Lowell, MA, USA), ferromagnetic resonance (FMR) spectrometer, and a magneto-optic Kerr effect (MOKE) microscope.[28] Magnetization versus applied magnetic field measurements were measured using VSM to determine saturation magnetization and to identify easy and hard axes of magnetization. Acquisition of a background for subtraction of the diamagnetic substrate signal was accomplished from measurement of an identical 5 mm x 5 mm substrate cleave without the magnetic multilayered film. FMR spectroscopy was conducted on a grounded coplanar waveguide (GCPWG), implementing a 40 GHz zero-bias diode detector (Krytar, Sunnyvale, CA, USA) and a 50 GHz *rf* current source (Agilent HP86530, Keysight, Santa Rosa CA, USA).[28] Using a fixed *rf* current amplitude ($I_{pk}^{rf} \cong 0.4$ mA) and frequencies between 10 GHz and 30 GHz, the magnetic field from a dc electromagnet was swept between 24 kA/m and 1000 kA/m. A pair of secondary coils embedded in the dc electromagnet provides a low frequency alternating field ($f = 177$ Hz, $\mu_0 H_{pk}^{ac} \cong 2$ mT) for lock-in detection of the differential power absorbed by the GCPWG and sample as the dc magnetic field is swept. The applied dc and ac fields were monitored with a Hall probe sensor. MOKE microscopy was conducted with a 10x magnifier and two polarizing lenses whose polarizations are 2-5 degrees away from orthogonal. Applied dc fields in the MOKE microscope were generated with an electromagnet, and were correlated back to currents in the electromagnet using a Hall probe sensor (Lake Shore Cryotronics, Westerville, OH, USA).[28]

**Author Contributions:**

D. G. and R. S. designed the study, in particular conceiving the strain-controlled magnetism. P. C. supported the synthesis of the magnetic multilayered films. C. D. and D. G.

collected and analysed the magnetic data. P. F. and M. S. directed the strain measurements. Y. I. supported the MOKE microscopy experiments. All authors contributed to the scientific process and the refinement of the manuscript.

**Additional Information:**

The authors declare no competing financial interests.

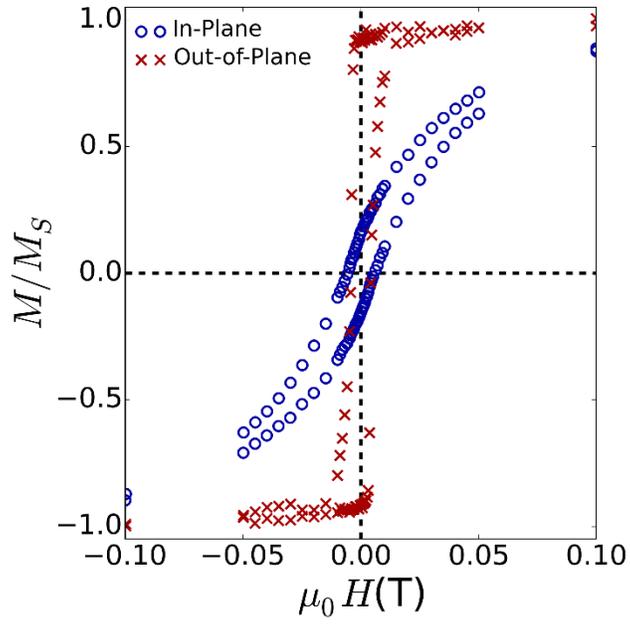

Figure 1: Normalized magnetization ($M/M_S$) versus applied magnetic field for the [Co(0.15)/Ni(0.6)]x16/Co(0.15) film showing out-of-plane easy axis. Magnetic fields are applied either in-plane (open blue circles) or out-of-plane (red crosses).

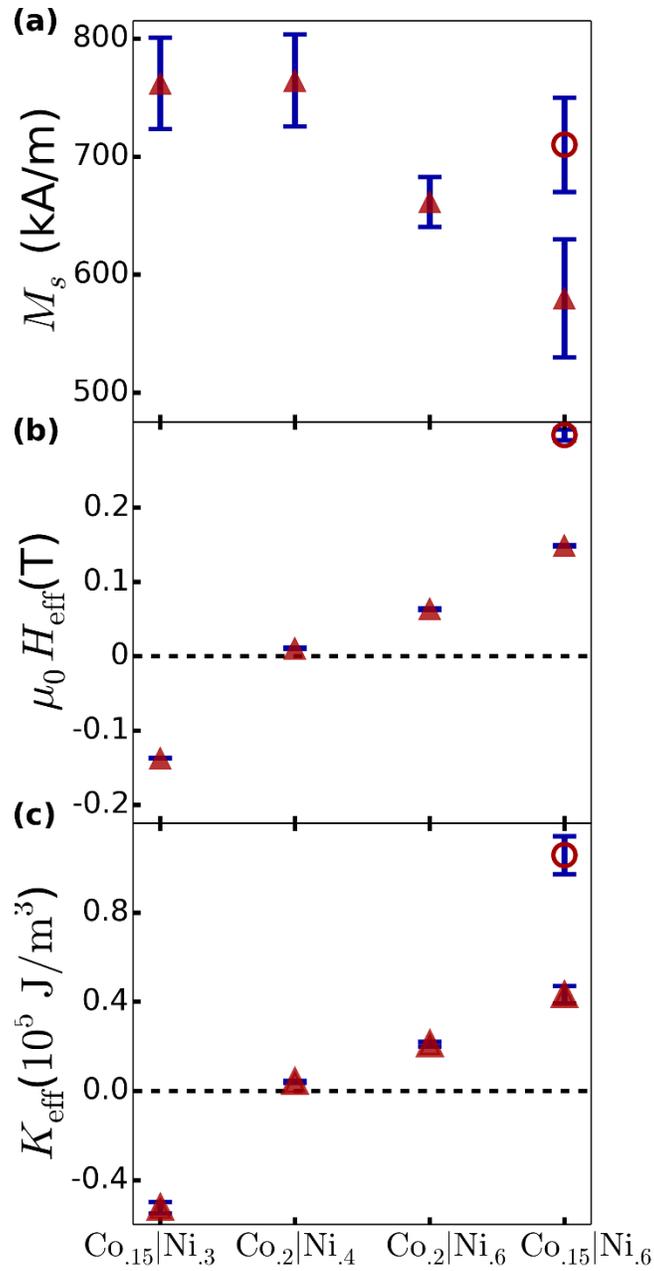

Figure 2: (a) Saturation magnetization determined from magnetization versus applied out-of-plane field hysteresis loops; (b) effective anisotropy field determined from ferromagnetic resonance frequency versus applied out-of-plane magnetic field and (c) anisotropy energy evaluated from (a) and (b) results. All solid markers represent samples containing 16 bilayer repeats of $Co_x|Ni_y$ (x,y are layer thicknesses in nm), with the unfilled round marker represents the sample consisting of 4 bilayers of $Co_{0.15}|Ni_{0.6}$.

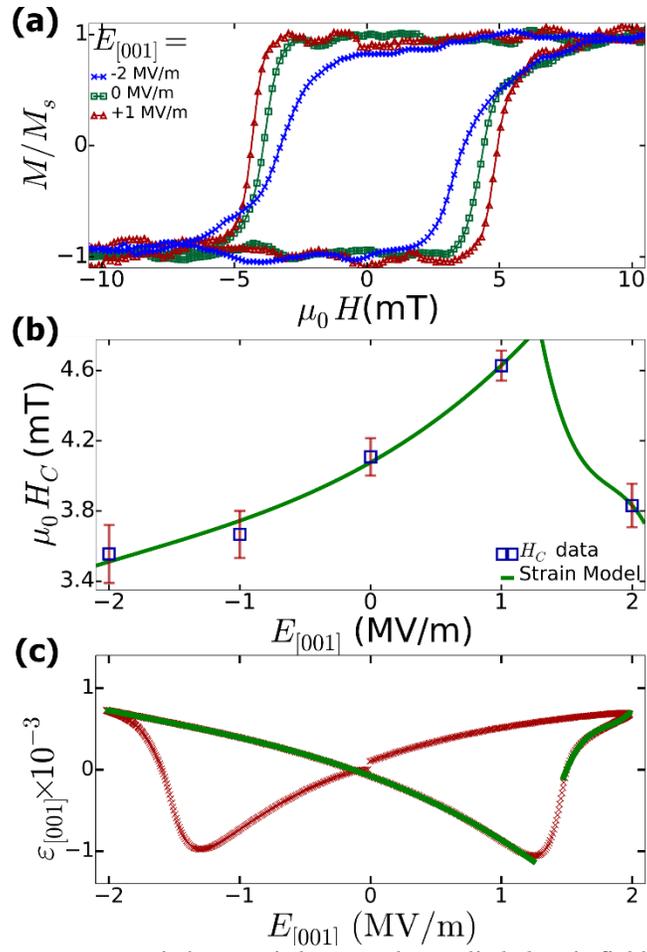

Figure 3: (a) Kerr microscopy magnetic hysteresis loops under applied electric fields to the 1 mm thick PZT substrate; (b) magnetic coercivity (blue open squares) versus applied electric field parameterized (green curve) by non-linear strain-electric field curve obtained for the PZT substrate in (c)

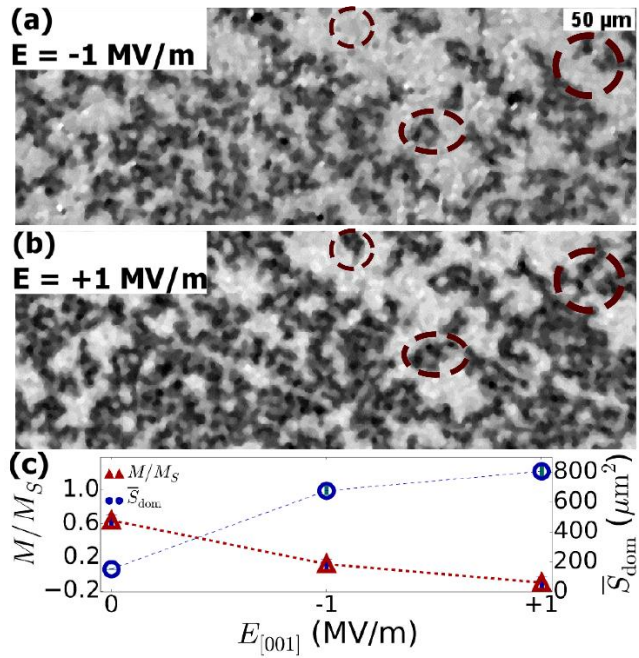

Figure 4: Kerr microscopy images of domain expansion under a modest out-of-plane magnetic field (-2.4 mT) and subsequent applications of electric fields to the PZT substrate of (a) -1 MV/m and (b) +1 MV/m (Circles in (a) and (b) highlight newly reversed areas due to electric field polarity switch) . Changes in the magnetization (red triangles) and the average reversed (dark contrast) domain size (open blue circles) are shown in (c) under the series of applied electric fields.

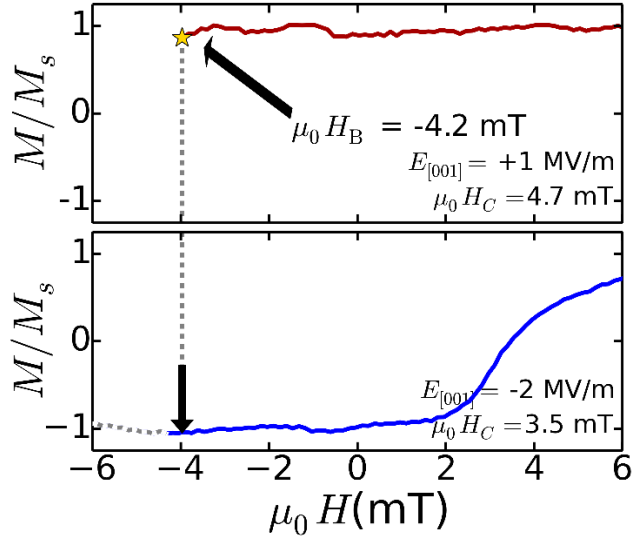

Figure 5: Concept illustrating voltage-assisted magnetization reversal. An out-of-plane bias magnetic field (-4.2 mT) is applied to a Co/Ni multilayered film on a PZT substrate under +1 MV/m electric field bias after saturation in a large positive magnetic field (red curve). Changing the electric bias to -2 MV/m drives the Co/Ni film into the low coercivity operating region, indicated by the arrow extending down from the yellow star. The Co/Ni film would subsequently follow the hysteresis curve defined by the low coercivity state (blue line).